\documentclass[12pt,preprint]{aastex}
\input psfig.sty
\usepackage{epsf}

\def\beq{\begin{equation}}
\def\eeq{\end{equation}}
\def\bey{\begin{eqnarray}}
\def\eey{\end{eqnarray}}
\def\pppm{\rm P^3M}
\def\mpc{\,h^{-1}{\rm {Mpc}}}
\def\mpci{\,h{\rm {Mpc}}^{-1}}

\def\kms{\,{\rm {km\, s^{-1}}}}
\def\msun{{h^{-1} M_\odot}}

\def\xiz#1{\xi_z(r_{p#1},\pi_{#1})}

\def\mag{M_b - 5 \log_{10} h}

\def\gs{\mathrel{\raise1.16pt\hbox{$>$}\kern-7.0pt
\lower3.06pt\hbox{{$\scriptstyle \sim$}}}}
\def\ls{\mathrel{\raise1.16pt\hbox{$<$}\kern-7.0pt
\lower3.06pt\hbox{{$\scriptstyle \sim$}}}}
\def\gtsima{$\; \buildrel > \over \sim \;$}
\def\ltsima{$\; \buildrel < \over \sim \;$}
\def\prosima{$\; \buildrel \propto \over \sim \;$}
\def\gsim{\lower.5ex\hbox{\gtsima}}
\def\lsim{\lower.5ex\hbox{\ltsima}}
\def\simgt{\lower.5ex\hbox{\gtsima}}
\def\simlt{\lower.5ex\hbox{\ltsima}}
\def\simpr{\lower.5ex\hbox{\prosima}}

\def\ga{\gsim}

\begin{document}
\title { The pairwise velocity dispersion of galaxies: luminosity
dependence and a new test of galaxy formation models}
\author{Y.P. Jing${^{1,2}}$, G. B\"orner${^{2,1}}$}
\affil{${^1}$Shanghai Astronomical Observatory, the Partner Group of
MPI f\"ur Astrophysik, \\Nandan Road 80, Shanghai 200030, China}
\affil {${^2}$Max-Planck-Institut f\"ur Astrophysik,
Karl-Schwarzschild-Strasse 1, \\ 85748 Garching, Germany}

\begin{abstract}
We present the first determination of the pairwise velocity dispersion
(PVD) for galaxies in different luminosity intervals using the final
release of the Two-Degree Field Galaxy Redshift Survey (2dFGRS). We
have discovered quite surprisingly that the relative velocities of the
faint galaxies at small separation are very high, around $700 \kms$,
reaching similar values as the brightest galaxies.  At intermediate
luminosities $M^*-1$ ( $M^*$ is the characteristic luminosity of the
Schechter function), the relative velocities exhibit a well defined
steep minimum near $400 \kms$.  This result has been derived using a
novel method to determine the real space power spectrum and the PVD
from the redshift space power spectrum of the 2dFGRS. Combined with
the observed luminosity dependence of clustering, our result implies
that quite a fraction of faint galaxies, as well as the brightest
ones, are in massive halos of galaxy cluster size, but most of the
$M^*$ galaxies are in galactic halos. Our observed result is compared
with the current halo model of galaxies of Yang et al. that was
obtained by matching the clustering and luminosity functions of the
2dFGRS. With the model parameters they favored most, the halo
model seems to be unable to reproduce the luminosity dependence of the
PVD because it predicts a monotonically increasing PVD with the
luminosity. We discuss a possible solution to this model by raising
the faint end slope of the conditional luminosity function in rich
clusters. The PVD luminosity dependence may also be an important
constraint in general on theories of galaxy formation, such as
semi-analytical models and hydro/N-body simulations of galaxy
formation.

\end{abstract}


\section {Introduction}

The clustering of galaxies in the Universe is characterized by their
spatial positions, and by their peculiar velocities that lead to
deviations of their motion from the pure Hubble flow.  The big
redshift surveys assembled in recent years by the diligent work of
many astronomers give angular positions and redshifts for large
numbers of galaxies. A rough 3D map can be obtained by placing the
galaxies at distances along the line of sight derived via Hubble's law
from their redshifts. The peculiar velocity, however, also contributes
to the redshift, and this leads to a misplacement of the galaxy away
from its true location.  The local gravitational field is the cause of
the peculiar motion, thus the redshift distortion in the galaxy
maps can give information on the underlying matter distribution.

The amplitude of the distortions can be estimated from the pair
distribution of galaxies.  For pairs of galaxies at distances much
larger than their separation, one can use a plane-parallel
approximation. In the linear approximation (or large separation), the
power spectrum of their distribution in redshift space is \citep{k87},
\beq 
P^S({\bf k}) = (1 + \beta \mu^2)^2 P(k) \,.
\label{eq1}
\eeq Here $\mu$ is the cosine of the angle between the wave vector and
the line of sight.  The linear redshift distortion parameter $\beta$
that is related to the linear growth factor $f(\Omega_0) \simeq
\Omega^{0.6}$ ($\Omega_0$ is the matter density) and the linear bias
factor $b$ of the galaxies by $\beta = \Omega_0^{0.6}/b$ , can be
estimated if $P^S({\bf k})$ can be measured on a sufficiently large
scale. $P(k)$ is the power spectrum in real space. The dependence on
$b$ expresses the fact that there is a bias in the galaxy
distribution, i.e. there are differences between the galaxy and the
dark matter distributions. The linear bias relation is just a constant
ratio of the galaxy $\xi$ and the dark matter $\xi_{dm}$ two-point
correlation functions (2PCF) $ \xi = b^2 \xi_{dm}$. It has been shown
that Eq.(\ref{eq1}) is valid only on sufficiently large scales
\citep[perhaps $\ga 20 \mpc$ or $k<0.1\mpci$,][]{scocci04}. On smaller
scales, the virial motion of galaxies within groups and clusters
contributes significantly to the deviation of the redshift distortion
from the prediction of Eq.(\ref{eq1}). In fact, on small scales ($\ls
5\mpc$), the virial motion of galaxies dominates the redshift
distortion.

Thus the redshift distortions on small scales can be used to
determine the pairwise peculiar velocity dispersion (PVD) of
galaxies. Assuming certain functional forms for the distribution
function of the pairwise velocity \citep[say, an exponential
form,][]{peebles76} and for the average infall velocity, a model can
be constructed for the redshift 2PCF which approximates the real
situation well, when the coupling between the peculiar velocity and
the spatial density of the galaxies is weak \citep[][hereafter
DP83]{fisher94, jfs98, dp83}. A comparison of the model with
observations of the redshift 2PCF provides a test for the validity of
the assumptions (such as the distribution function of the pairwise
velocity) and a determination of the
PVD $\sigma_{12}(r)$. DP83 applied this method to the CfA redshift
survey, and determined the PVD to be $340\pm 40\kms$ at projected
separation $r_p = 1\mpc$. Based on an extensive study of the PVD for
all redshift surveys (typically containing 2000 galaxies) available
before 1993, \citet{mjb93} showed that the PVD measured is very
sensitive to the presence of rich clusters in the survey. Because the
CfA survey used by DP83 is too small to fairly represent the
population of rich clusters in the Universe, Mo et al. pointed out that
the PVD estimate given by DP83 is likely to be significantly biased
low based on their analysis of the PVD for different
surveys. Subsequent analyses for CfA2 by \citet{marzke95} and for CfA
by \citet{sdp97} have confirmed the conclusion of Mo et al..

With the Las Campanas redshift survey (LCRS) of galaxies,
\citet[][hereafter JMB98]{jmb98} made the first accurate determination
of the PVD using the above method. They used mock catalogs generated
from N-body simulations of the concordance $\Lambda$ cold dark matter
(LCDM) model to correct for the observational effects including the
fiber-collisions, and to make a fair estimation of errors for the
measured PVD. They demonstrated that the errors of their measured PVD
are between 50 and 100 $\kms$, and claimed that the PVD can be
accurately measured with the LCRS, in contrast with all previous work
based on smaller surveys. The PVD at the projected separation $1\mpc$
is $570\pm 80\kms$. This result has been verified by \citet{zehavi02}
with the early data release of the Sloan Digital Sky Survey(SDSS).
For IRAS galaxies, the PVD is lower than that of optical galaxies
\citep[]{fisher95,fisher94,mjb93}, and the PVD is very low on small
scales \citep[]{jbs02, hamilton02}. These results of IRAS galaxies are
also consistent with the SDSS study \citep[]{zehavi02} where it was
found that the PVD of the blue (young) galaxies is as low as that of
IRAS galaxies. Cosmological hydrodynamic simulations of galaxy
formation seem to lead to similar conclusions with regard to the PVD
of different galaxy populations
\citep[]{pearce01,yoshikawa03,berlind03,wdkh04}.

Another way to measure the PVD is to use the redshift space power
spectrum.  The redshift power spectrum $P^S(k,\mu)$ can be written as
\citep{pd94,cfw95}
\beq
P^S(k.\mu) = P(k)(1+\beta \mu^2)^2 D(k \mu \sigma_{12}(k))\,
\label{eq2}
\eeq
where the first term is the Kaiser linear compression, and the second
term $D$ is the damping effect caused by the random motion of the galaxies. For
the exponential distribution function of the pairwise velocity, the
function $D$ is Lorentz form
\beq
D(k \mu \sigma_{12}(k)) = \frac{1}{1+ \frac{1}{2}k^2 \mu^2 \sigma_{12}(k)^2}\,.
\label{eq3}
\eeq 
\citet{jb01} have studied the redshift power spectrum in typical
CDM models for different tracers --the primordial density peaks model,
the cluster-underweighted scheme of particles, and pure dark matter,
and found that different tracers may have different forms of the
damping function. Thus, the damping function can serve as a 
constraint on galaxy formation models. From the analysis of the LCRS
catalog, both \citet{lsb98} and \citet{jb01b} found that the damping
function of observed galaxies is very close to
Eq.(\ref{eq3}). Adopting Eq.(3) for the damping function,
\citet{jb01b} measured the PVD
$\sigma_{12}(k)$ for LCRS by setting $\beta=0.45$ and found that the
measured PVD is consistent with the PVD that JMB98 reported on the
redshift 2PCF measurement. Hawkins et al. (2003) used Eq.(2) and
Eq.(3) to model the redshift 2PCF with the Fourier transform, and
measured the $\beta$ and $\sigma_{12}$ (assuming
$\sigma_{12}(k)=const.$) for the Two-degree Field Galaxy Redshift
Survey (2dFGRS) by best fitting the observed 2PCF. The advantage of
using the redshift power spectrum to determine the PVD is that it is
simple and accurate to model the infall effect.

Here we use the data of the final release of the
2dFGRS\footnote{Available at http://www.mso.anu.edu.au/2dFGRS} for
such an analysis to study the luminosity dependence of the PVD. The
parameter $\beta$ could, however, not be determined independently for
each luminosity subsample, since there are still large statistical
fluctuations on large scales $ k \sigma \mu \leq 1$.  Therefore, we
fix $\beta$ at a reasonable value of $ 0.45$. Later we come back and
consider the influence of a luminosity dependence \citep[as
in][]{norberg02a} of $\beta$.  Inspection of
Fig.~(\ref{fig:sigmabetalum}) shows that the results are basically
unchanged, implying that the results are robust to reasonable changes
of the $\beta$ values.

The 2dFGRS has already been analysed statistically with respect to the
PVD \citep[]{hawkins03} and the determination of $\beta$
\citep[]{peacock00,hawkins03}. But here we use a novel method to
estimate the PVD. Furthermore, we take advantage of the large number
of galaxies in the 2dFGRS to bin the galaxies in different luminosity
intervals, and study the luminosity dependence of the PVD. This was
not possible up to now, and we shall see that remarkable tests of
galaxy formation models become possible with the statistical results
presented here.  The luminosity dependence of the clustering (the
2PCF) of galaxies in the 2dFGRS has been investigated
\citep[]{norberg01, norberg02a, norberg02b} but not of the PVD.

It has been demonstrated already in an analysis of the LCRS (JMB98)
that the observationally measured 2PCF and the PVD require a scale
dependent bias model in order to find acceptable theoretical fits. The
so-called cluster-underweighted model that was invented by JMB98 to
bring a flat $\Omega = 0.2$ CDM model into agreement with the data has
meanwhile been developed into a more sophisticated
``halo-occupation-distribution'' (HOD) model
\citep[e.g.][]{sheth01,bw02,kang02}. In particular,
\citet[][hereafter Y03]{yang03} introduced the conditional luminosity
function $\Phi(L\vert M)$ to characterize the luminosity distribution
of galaxies within a halo of mass $M$, and measured the parametrized
form of $\Phi(L\vert M)$ by matching the observed luminosity function
of galaxies and the luminosity dependence of galaxy clustering in the
2dFGRS. This model has been applied to predict the clustering and
velocity statistics of galaxies in the DEEP2 survey \citep{ymw03}. By
construction, the model of Y03 is able to reproduce the luminosity
function and the two-point correlation function of galaxies. But the
PVD, especially its luminosity dependence, can serve as an independent
test on the model, as the PVD on small scales is rather sensitive to
the HOD.  We will show that the model most favored by Y03 is unable
to match our measured luminosity dependence of the PVD, indicating
that some of the parameterizations adopted by Y03 have to be amended.
A comparison with the semi-analytic models for galaxy formation
\citep[]{dkcw99, benson00} will be presented in a subsequent
investigation.

In the following chapters we describe our way of using the redshift
power spectrum, the construction of mock catalogs from our
simulations, tests of the method, the results, and the comparison with
the HOD model of Y03. In the final discussion we summarize
our results, and suggest construction principles for better models.

\section{Observational sample and random sample}
We select data for our analysis from the final release of the 2dFGRS
 \citep[][hereafter C01]{colless03, colless01}. The survey covers two
 declination strips, one in the Southern Galactic Pole (SGP) and
 another in the Northern Galactic Pole(NGP), and 99 random fields in
 the southern galactic cap. Each redshift determination is assigned a
 quality class $Q$ in the 5-point system according to the measurement
 accuracy based on emission and absorption lines. Quality $Q=1$ or 2
 means a doubtful redshift estimate, $Q=3$ means a probable redshift
 with the confidence 90 percent, and $Q=4$ or $5$ means a reliable
 redshift. Quality classes 1 and 2 are considered failures. The
 redshift sampling completeness $R({\mathbf \theta})$ (${\mathbf
 \theta}$ is a sky position) that is defined as the fraction of
 targeted galaxies for which a redshift is measured with $Q\ge 3$, is
 available for each sky sector (C01). In this paper, only galaxies in
 the two strips are considered. Further criteria for galaxies to be
 included in our analysis are that they are within the redshift range
 of $0.02<z<0.25$, have the redshift measurement quality $Q\ge 3$, and
 are in the regions with the redshift sampling completeness
 $R({\mathbf \theta})$ better than 0.1 (where ${\mathbf \theta}$ is a
 sky position).  The redshift range restriction should ensure that the
 clustering statistics are less affected by the galaxies in the local
 supercluster, and by the sparse sampling at high redshift. The
 redshift quality restriction is imposed so that only galaxies with
 reliable redshift are used in our analysis. An additional reason is
 that the redshift completeness mask provided by the survey team,
 which is used in our analysis, is constructed for the redshift
 catalog of $Q\ge 3$. The last restriction is imposed in order to
 eliminate galaxies in the fields for which the field redshift
 completeness $c_{F}$ is less than 70 percent (see C01 about the
 difference between $R({\mathbf \theta})$ and $c_F$). These fields
 have not been included in computing the redshift mask map $R({\mathbf
 \theta})$. As a result, there are a total of 190504 galaxies
 satisfying our selection criteria, 78190 in the NGP strip and 112314
 in the SGP strip.

In order to study the luminosity dependence of the PVD, we divide the
galaxies into 10 subsamples according to their absolute
magnitude. The subsamples are successively brightened by 0.5
magnitude from the faintest sample $M_b=-17.0+5\log_{10} h$ to
$M_b=-21.5+5\log_{10} h$, with successive subsamples overlapping by
0.5 magnitude. Here $h$ is the Hubble constant in units of
$100\kms{\rm Mpc}^{-1}$.  The details of the subsamples studied in
this paper are given in Table 1.  For computing the absolute
magnitude, we have used the k-correction and luminosity evolution
model of \citet[][ ${\rm k+e}$ model]{norberg02b}, i.e., the absolute
magnitude is in the rest frame $b_j$ band at $z=0$. We assume a
cosmological model with the density parameter $\Omega_0=0.3$ and the
cosmological constant $\lambda_0=0.7$ throughout this paper.

A detailed account of the observational selection effects is released
with the catalog by the survey team (C01). The limiting magnitude
changes slightly across the survey region due to further magnitude
calibrations that were carried out after the target galaxies had been
selected for the redshift measurement. This observational effect is
documented in the magnitude limit mask $b_J^{\rm lim} (\mathbf \theta)
$ (C01). The redshift sampling is far from uniform within the survey
region, and this selection effect is given by the redshift
completeness mask $R (\mathbf \theta)$. The redshift measurement
success rate also depends on the brightness of galaxies, making
fainter galaxies more incomplete in the redshift measurement. The $\mu
( \mathbf \theta)$ mask provided by the survey team is aimed to
account for the brightness-dependent incompleteness. These effects can
be corrected in computing the redshift two-point correlation function
through constructing random samples that properly include these
selection effects. We generate random samples in the same way as
described in \citet[]{jb04}. Each random sample for a northern or
southern luminosity subsample contains 100,000 random points.

\section{Redshift power spectrum and the pairwise velocity dispersion}

We measure the redshift two-point correlation functions $\xi_z({\bf s})$ 
following the method of JMB98.

We convert the redshift two-point correlation function  
to the redshift power spectrum by the Fourier transformation:
\beq
P^S({\bf k})=\int \xi_z({\bf s}) e^{-i{\bf k}\cdot {\bf s}} d{\bf s}\,.
\label{eq31}
\eeq
In cylindrical polar coordinates ($r_p, \phi, \pi$) with the
$\pi$-axis parallel to the line-of-sight, $P^S({\bf k})$ depends on
$k_p$, the wavenumber perpendicular to the line-of-sight, and on
$k_\pi$, the wavenumber parallel to the line-of-sight. The power spectrum
can be written
\beq
P^S(k_p,k_{\pi})=\int \xiz{} e^{-i[k_p r_p\cos(\phi)+\pi k_\pi]}
r_p dr_p d\phi d\pi\,.
\label{eq4}
\eeq
With some elementary mathematical manipulation, we get the following
expression:
\beq
P^S(k_p,k_{\pi})=2\pi \int_{-\infty}^{\infty} d\pi \int_0^{\infty} r_p dr_p
\xiz{} \cos(k_\pi \pi) J_0(k_p r_p)
\label{eq5}
\eeq
where $J_0(k_p r_p)$ is the zeroth-order Bessel function \citep[]{jb01b}.

$\xiz{}$ is measured in equal logarithmic bins of
$r_p$ and in equal linear bins of $\pi$. The reason why different
types of bins are chosen for $r_p$ and $\pi$ is the fact that $\xiz{}$
decreases rapidly with $r_p$ but is flat with $\pi$ on small
scales. Thus this way of presenting $\xiz{}$ is better than using the
$\log$-$\log$ or the linear-linear bins for $r_p$ and $\pi$, and is
also suitable for the present work.  The peculiar velocity of a few
hundred $\kms$ should smoothen out structures on a few $\mpc$ in the
radial direction, and the linear bin of $\Delta \pi_{i}=1\mpc$ is
suitable for resolving the structures in the radial direction. With
logarithmic bins chosen for $r_p$, the $r_p$ dependence is resolved
well, because otherwise the small scale clustering on the projected
direction cannot be recovered.
With this bin method, we obtain the power spectrum:
\beq
P^S(k_p,k_{\pi})=2\pi \sum_{i,j} \Delta \pi_{i} r_{p,j}^2 \Delta \ln r_{p,j}
\xi_z(r_{p,j},\pi_i) \cos(k_\pi \pi_i) J_0(k_p r_{p,j})
\label{eq6}
\eeq where $\pi_{i}$ runs from $-50$ to $50\mpc$ with $\Delta
\pi_{i}=1\mpc$ and $r_{p,j}$ from $0.1$ to $50\mpc$ with $\Delta \ln
r_{p,j}=0.23$ (Be careful not to confuse two $\pi$s in
Eqs. (\ref{eq5}) and (\ref{eq6}): the first $\pi$ in the
right-hand-side has the conventional meaning, i.e. 3.14159..., and the
others are for the axis along the line-of-sight.). We make the
summation of Eq.(\ref{eq6}) with rectangular boundaries in $\pi$ and
$r_p$.

The fluctuations of $\xiz{}$ at large separations may bring errors to
the determination of the redshift power spectrum. We improve the
measurement by down weighting $\xiz{}$ at the larger
scales. Specifically, we use a Gaussian window function
\beq
W_g({\bf s})=\exp (-\frac{1}{2} \frac{s^2}{S^2})
\eeq
to weight the two-point correlation function with $S=20\mpc$. That
is, the measured redshift power spectrum is
\beq
P^{S,m} (k_p,k_{\pi})=2\pi \sum_{i,j} \Delta \pi_{i} r_{p,j}^2 \Delta 
\ln r_{p,j}
\xi_z(r_{p,j},\pi_i) W_g(r_{p,j},\pi_i) \cos(k_\pi \pi_i) J_0(k_p r_{p,j})\,.
\label{psweight}
\eeq 

The weighting reduces the noise in the measurement, but introduces a
systematic bias to the redshift power spectrum. This is the reason
why we distinguish this measured redshift power spectrum $P^{S,m}
(k_p,k_{\pi})$ from the expression $P^{S} (k_p,k_{\pi})$ in
Eq.(\ref{eq6}). But they are related by the following equation,
\beq
P^{S,m} ({\mathbf k})=\frac{1}{(2\pi)^3} \int  P^{S} ({\mathbf k}_1)
W_g({\mathbf k}-{\mathbf k}_1; S) d {\mathbf k}_1
\label{psconv}
\eeq
and
\beq
W_g({\mathbf k}; S)=\frac{1}{(2\pi)^{3/2}S^3} \exp (-\frac{1}{2}S^2k^2)
\,.
\eeq

\begin{figure}
\epsscale{1.0} \plotone{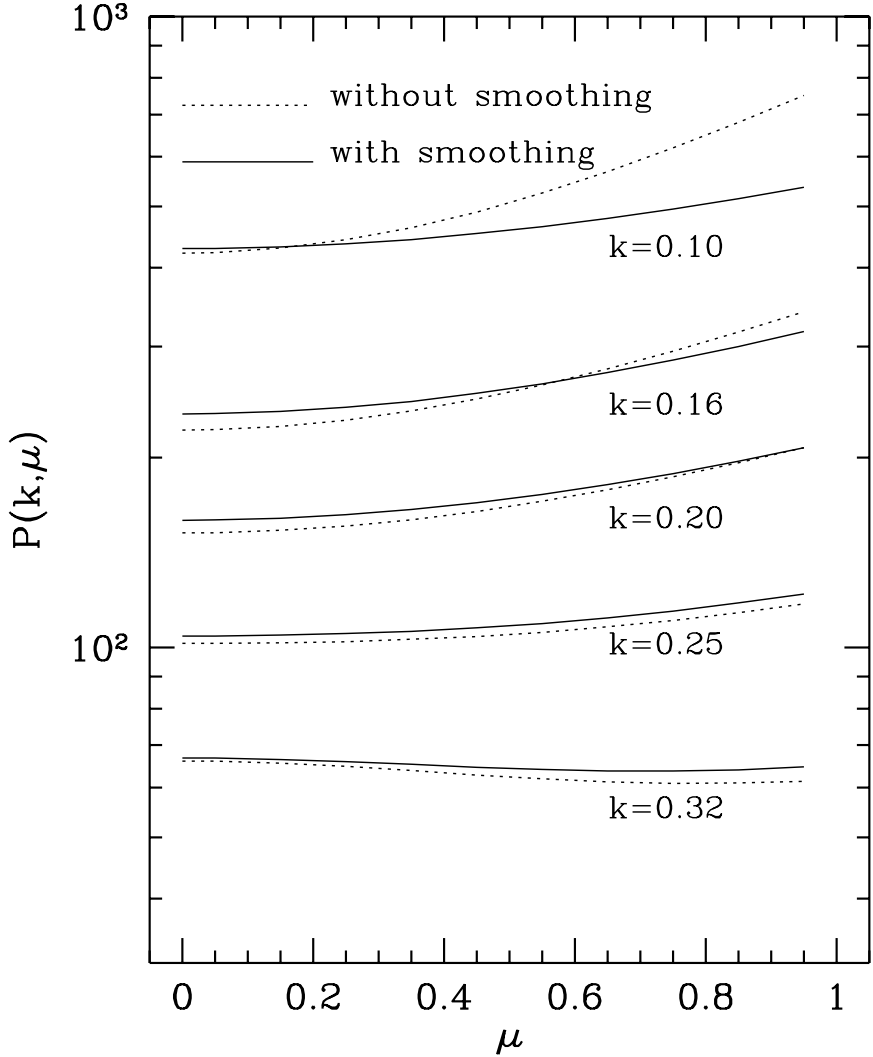}
\caption{Illustration of the smoothing effect on the determination of
the redshift space power spectrum $P(k,\mu)$ when the redshift space
two-point correlation function $\xi(r_p,\pi)$ is weighted by a
Gaussian function of width $R=20 \mpc$. The solid lines, computed with
Eq.(\ref{psconv}), are the $P(k,\mu)$ with the smoothing effect,
compared to the dotted lines for $P(k,\mu)$ without the smoothing
effect. In this plot, we use the linear CDM power spectrum of
$\Gamma=0.2$, $\beta=0.45$ and $\sigma_v=500 \kms$ as input. $k$ is in
units of $\mpci$}
\label{fig:smoothing}\end{figure}

The weighting may change the redshift power spectrum at $k\ls
\pi/S=0.15 \mpci$, making the spectrum more isotropic and biasing the
value of $P^S(k_p,k_{\pi})$ generally. This effect can be estimated as
follows. We take Eqs.(\ref{eq2}) and (\ref{eq3}) for the redshift
power spectrum $P^S({\mathbf k})$. As a typical example, we take
$\beta=0.45$, $\sigma_v=500 \kms$, and a linear Cold Dark Matter power
spectrum from \citet[]{bardeen86} with the shape parameter
$\Gamma=0.2$ for $P^S({\mathbf k})$, and compute $P^{S,m} ({\mathbf
k})$ with Eq.(\ref{psconv}). The result is plotted in
Fig.~(\ref{fig:smoothing}) which is also compared with $P^{S}
({\mathbf k})$. The weighting makes the redshift power spectrum
significantly rounder at $k= 0.1 \mpci$, but the effect becomes
negligible for $k \gs 0.2\mpci$. Furthermore, the figure shows that
the $P^{S} (k,\mu=0)$ at $\mu=0$ is changed little even at
$k=0.1\mpci$, indicating that the real space power spectrum may be
measured unbiasedly for $k \ge 0.1\mpci$ (because $P(k)=P^{S} (k,
\mu=0)$).

\begin{figure}
\epsscale{1.0} \plotone{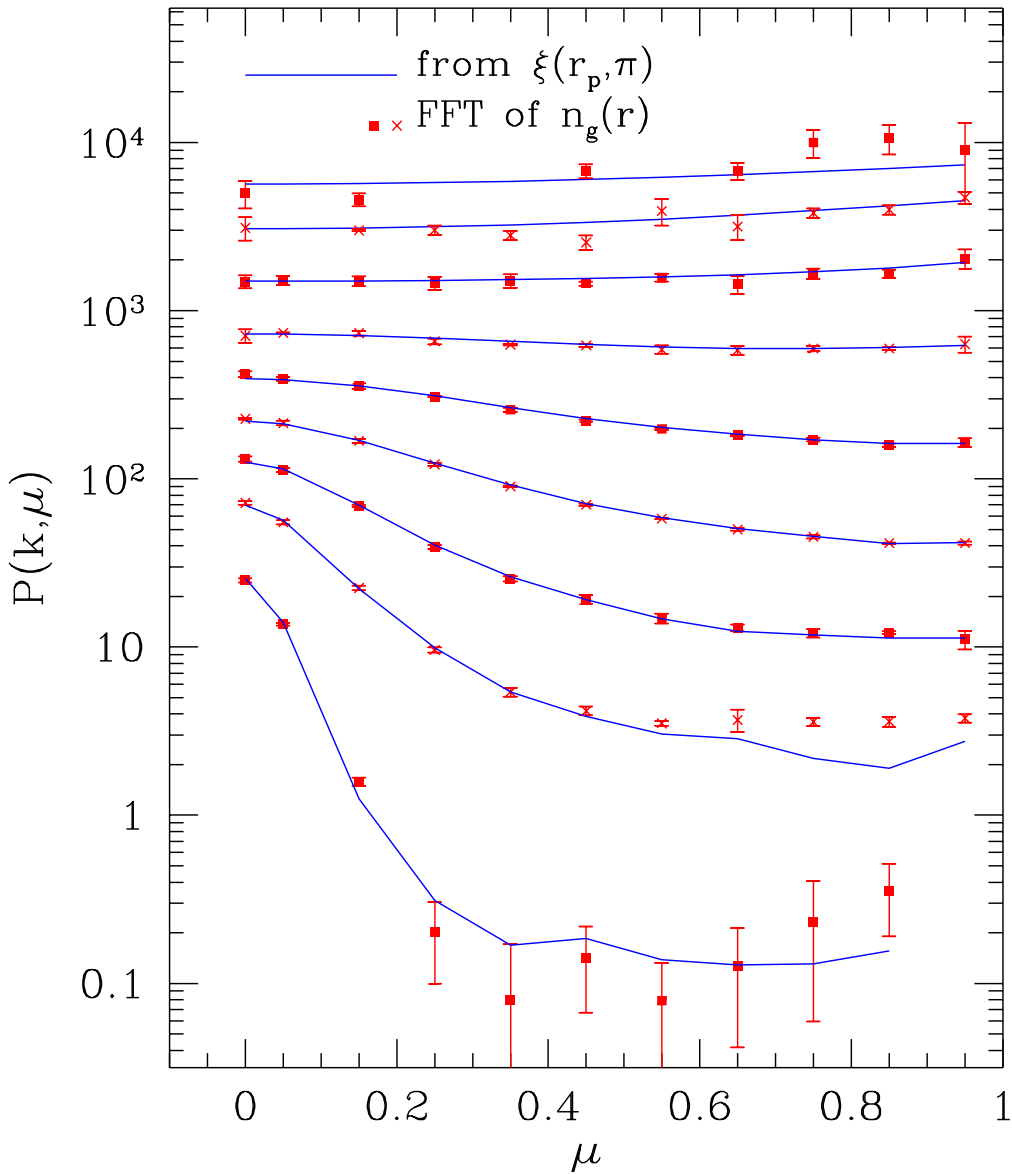}
\caption{The simulation test of the statistical method used in this
paper that measures $P(k,\mu)$ from $\xi(r_p,\pi)$. The solid lines
are the measurement of $P(k,\mu)$ for a simulation sample of galaxies
with $-19.5< M_b-5\log_{10} h <-18.5$ based on the Gaussian weighted
$\xi(r_p,\pi)$ measurement, compared with the direct measurement of
$P(k,\mu)$ based on the Fourier transformation of the galaxy density
field in redshift space (the symbols). From top to bottom, the
wavelength $k$ is $0.10, 0.16, 0.25, 0.40, 0.63, 1.0, 1.6, 2.5,$ and
$5.0\mpci$ respectively. The galaxies are produced with the halo
model, and the simulations have a boxsize $L=300\mpc$.  }
\label{fig:simutest}\end{figure}

A further test of the method is carried out by comparing the redshift
power spectrum of a simulated galaxy catalog. The galaxy catalog is
generated using the halo model \citep[Y03,][hereafter Y04]{yang04} for a LCDM
simulation of boxsize $300\mpc$. The details about the CDM model, the
simulation, and the method of generating the galaxies will be given in
the next section. For our test, we select galaxies with luminosity in
$-19.5\le M_b-\log_{10} h <-18.5$. We computed the redshift two-point
correlation function by counting the galaxy pairs, and measure the
redshift power spectrum using the procedure outlined above. These
spectra are compared with the redshift power spectra measured with the
Fourier transformation (FT) of the galaxy density field. This is shown
in Fig.~(\ref{fig:simutest}), with the symbols for $P(k,\mu)$ obtained
with FT, and the line for those measured through the 2PCF.  The power
spectra measured with FT are free of the weighting effect
[Eq.(\ref{psconv})]. The figure confirms the results of
Fig.~(\ref{fig:smoothing}), and indicates that our method can give an
unbiased estimate of $P(k,\mu)$ for $k\ge 0.2\mpci$ from which we will
measure the velocity dispersion $\sigma_v(k)$ through equation
(\ref{eq2}).  The real space power spectrum $P(k)$, which is mainly
determined from $P^S(k,\mu=0)$, can be determined for $k\ge 0.1\mpci$.

\section{Simulation, Halo model, and mock catalogs}
We simulate galaxy catalogs using our cosmological N-body simulations.
The cosmological model considered is a currently popular flat
low-density model with the density parameter $\Omega_0=0.3$ and the
cosmological constant $\lambda_0=0.7$ (LCDM). The shape parameter
$\Gamma=\Omega_0 h$ and the amplitude $\sigma_8$ of the linear density
power spectrum are 0.2 and 0.9 respectively. Two sets of simulations,
with boxsizes $L=100\mpc$ and $L=300\mpc$, that were generated with
our vectorized-parallel $\pppm$ code \citep[]{js02,jing02}, are used in this
paper. Both simulations use $512^3$ particles, so the particle mass
$m_p$ is $6.2\times 10^8\msun$ and $1.7\times 10^{10}\msun$
respectively in these two cases. We have four independent realizations
for each boxsize. Dark matter halos are identified with the
friends-of-friends method (FOF) using a linking length $b$ equal to
0.2 times of the mean particle separation. All halos with ten members
or more are included for generating the galaxy catalog.

\begin{figure}
\epsscale{1.0} \plotone{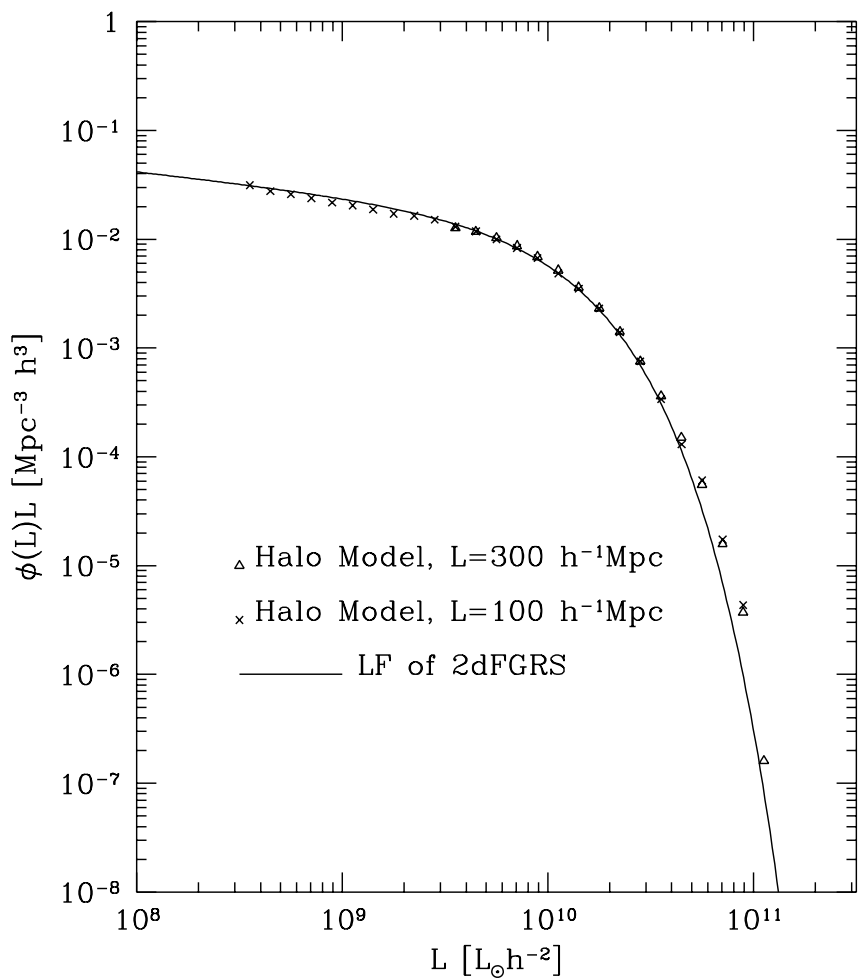}
\caption{The luminosity function of galaxies generated with the halo
model, compared with the observation of 2dFGRS. To the resolution
limit $M_b-5\log_{10} h =-18.5$ for the simulation of $L=300\mpc$ and
$M_b-5\log_{10} h =-16.5$ for the simulation of $L=100\mpc$, the luminosity
functions of the mock galaxies agree well with the observed one, and
agree with each other in the two simulations.}
\label{fig:lf}\end{figure}

We populate the halos with galaxies according to the prescription
proposed by Y03. The luminosity function of galaxies in a
halo is assumed to be a function of the halo mass $M$, $\Phi(L\vert
M)$ that is further parametrized as the Schechter function. The
parameters of the Schechter function are functions of $M$, further
parametrized through the mass-to-light ratios of halos. There are a
total of 8 parameters (without classifying galaxies with spectral
types) that are determined by best fitting the observed luminosity
function and the luminosity dependent clustering of galaxies in the
2dFGRS. It is shown by Yang et al. that these two observations are
well reproduced by their model, but there is some degeneracy in the
model parameters. In our work, we adopt model M1 in Y03
for the parameters of the halo model, and populate the halos with
galaxies (the luminosity, position and velocity) in a similar way as
Y04. The code for populating the halos
with galaxies was, however, written by us independently. We adopt the
``FOF satellites'' scheme of Y04. We assign the satellite galaxies of
a halo the position and velocity of dark matter particles randomly
selected from the halo. Instead of locating the central galaxy at the
center-of-mass of a halo as in Y04, we locate it at the potential
minimum of the halo and assign its velocity with the halo bulk
velocity.  The mass resolution of the simulations sets a faint limit
for which galaxies can be presented. According to Y04,
galaxies are complete in the $300\mpc$ simulation for magnitude
brighter than $M_b-5\log_{10} h=-18.0$ and in the $100\mpc$ simulation
for magnitude brighter than $M_b-5\log_{10} h=-15.0$. In
Fig.~(\ref{fig:lf}), we present the luminosity functions of the model
galaxies which do agree well with the observation of the 2dFGRS, and
also agree with each other between the two sets of simulations.

It is straightforward to produce mock catalogs of the 2dFGRS, since we
have the information of positions, velocity and luminosity for each
galaxy. We first duplicate the simulation volumes periodically along
the main axes, and execute the selection effects according to the
observational masks. We produce 5 mock galaxy catalogs for each
realization of a simulation, so we have a total 40 mock samples. For
studying the redshift power spectrum of galaxies fainter than
$M_b-5\log_{10} h=-18.0$, we will use the mock samples of the $100\mpc$
simulations, otherwise we will use those of the $300\mpc$ simulations.

\begin{figure}
\epsscale{1.0} \plotone{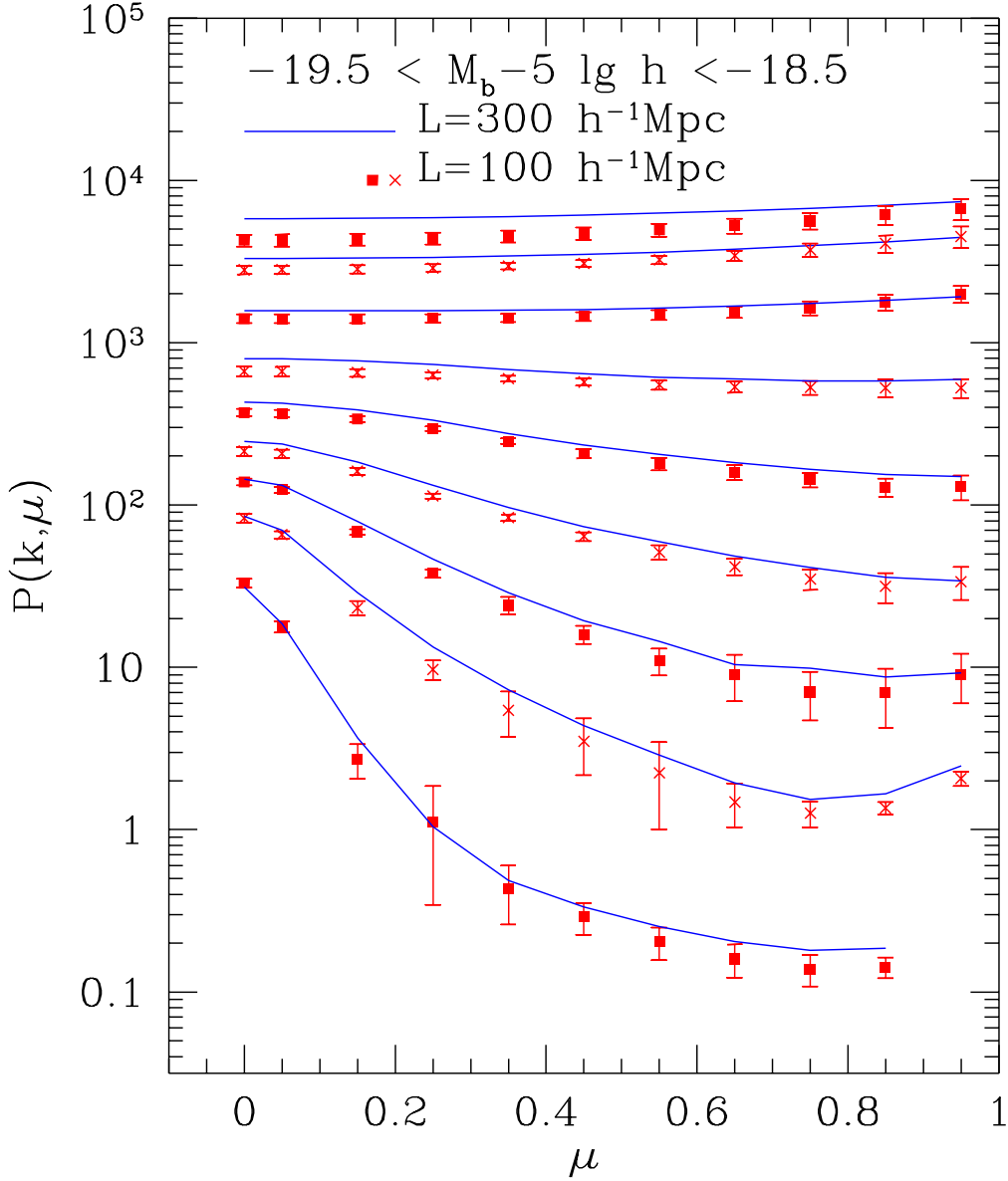}
\caption{The resolution effect on the predicted $P(k,\mu)$ in
simulations of different boxsizes. Galaxies of $-19.5< M_b-5\log_{10} h
<-18.5$ are analyzed. The results based on the 2dFGRS mock samples
generated with $L=100\mpc$ simulations are plotted in symbols and
those with $L=300\mpc$ simulations are plotted in connected
lines. From top to bottom, the wavelength $k$ is $0.10, 0.16, 0.25, 0.40, 0.63, 1.0, 1.6, 2.5$, and $5.0 \mpci$
respectively.}
\label{fig:resolution}\end{figure}

It is important to understand the numerical artifacts of the
simulations that may affect the result here. We believe that the force
resolution of the simulations that adopt $\pppm$, are sufficient for
the current study, since only the global information of halos is
needed. The mass resolution has already been discussed above. It can
be easily taken into account in our analysis. The boxsize may be the
only concern that we need to consider seriously, because the most
massive halos may be under-represented for the lack of long-wavelength
density fluctuations, and the redshift power spectrum cannot be
measured around the fundamental wavelength of the simulation. Since we
measure $P(k,\mu)$ for $k\ge 0.1\mpci$, we believe that the boxsize
$300\mpc$ is sufficiently large both for studying the large-scale
clustering and for fairly sampling the massive clusters. To quantify
the possible effect of the 100$\mpc$ boxsize, we measure the redshift
power spectrum for the luminosity $-19.0 < M_b-5\log_{10} h<-18.5$ from the
mock catalogs. The results of the two sets of mock catalogs are
compared in Fig.~(\ref{fig:resolution}), which clearly show that the
simulation of 100$\mpc$ box size is sufficiently big in volume for studying the
redshift power spectrum for $k > 0.16\mpci$. Also the population of the
halos that host the faint galaxies in the simulation of $100\mpc$
should not be affected by the limited volume.

\section{Results}
The statistical quantities are defined in the Fourier space ($\mathbf
k$) in the present work, which are often compared those obtained in
the coordinate space ($\mathbf r$) in previous work. When doing this
comparison, one uncertainty is the scale correspondence between the
$\mathbf k$ and $\mathbf r$ spaces. Although the correspondence is
$r=2\pi/k$ in mathematics, we think that it is more appropriate to
compare the two-point clustering and the PVD with the relation
$r\Leftrightarrow 1/k$ based on the following facts. The two-point
clustering can be expressed as a sum of two-halo and one-halo
contributions \citep[e.g.][]{kang02}. The transition from the two-halo
term dominance at large scales to the one-halo term dominance happens
at $\sim 0.5\mpci$ in $k$-space or at $r\sim 2\mpc$ in the $r$-space
for the concordance LCDM model\citep[e.g.][Y04]{kang02}. When
comparing the scale dependence of PVD, \citet{jb01} also argued for $k
\Leftrightarrow 1/r$. Therefore, we adopt $r\Leftrightarrow 1/k$ in
the current paper when comparing the statistical quantities in the two
spaces.

\subsection{The luminosity dependence of the PVD}
\begin{figure}
\epsscale{1.0} \plotone{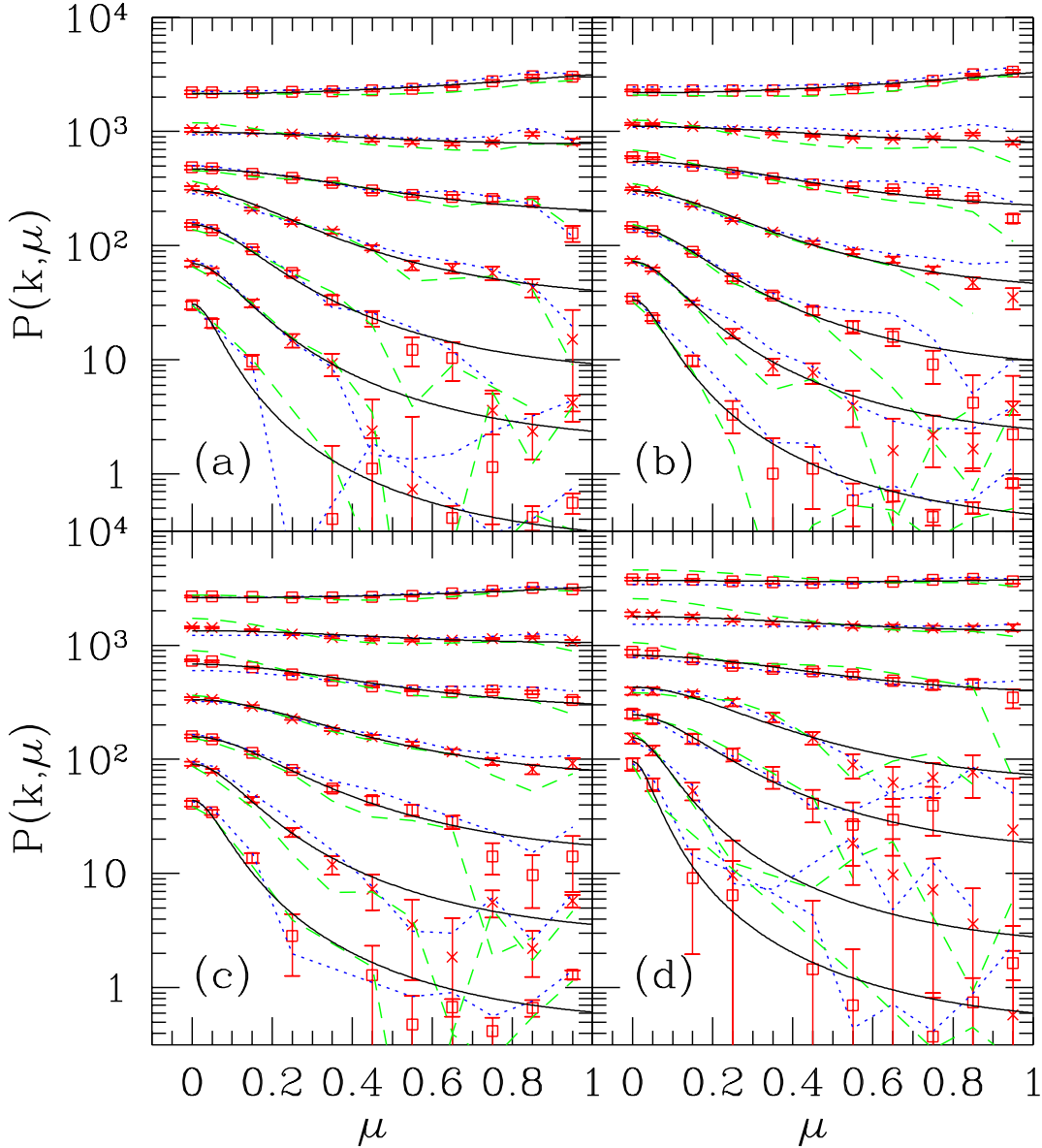}
\caption{The redshift space power spectrum $P(k,\mu)$ measured in
2dFGRS. The symbols are for the whole survey, the dotted lines for the
south subsample, and the dashed lines for the north subsample. The
errors are plotted only for the whole survey that is estimated with the
bootstrap method. The smooth solid lines are the best fits of
Eq.(\ref{eq2}) to data of the whole sample. (a) for $-18.5<
M_b-5\log_{10} h <-17.5$; (b) for $-19.5< M_b-5\log_{10} h <-18.5$;
(c) for $-20.5< M_b-5\log_{10} h <-19.5$; (d) for $-21.5<
M_b-5\log_{10} h <-20.5$.  In each panel from top to bottom, the
wavelength $k$ is $0.2, 0.32, 0.50, 0.79, 1.26, 2.0$, and $3.2\mpci$
respectively.}
\label{fig:pkmu}\end{figure}

\begin{figure}
\epsscale{1.0} \plotone{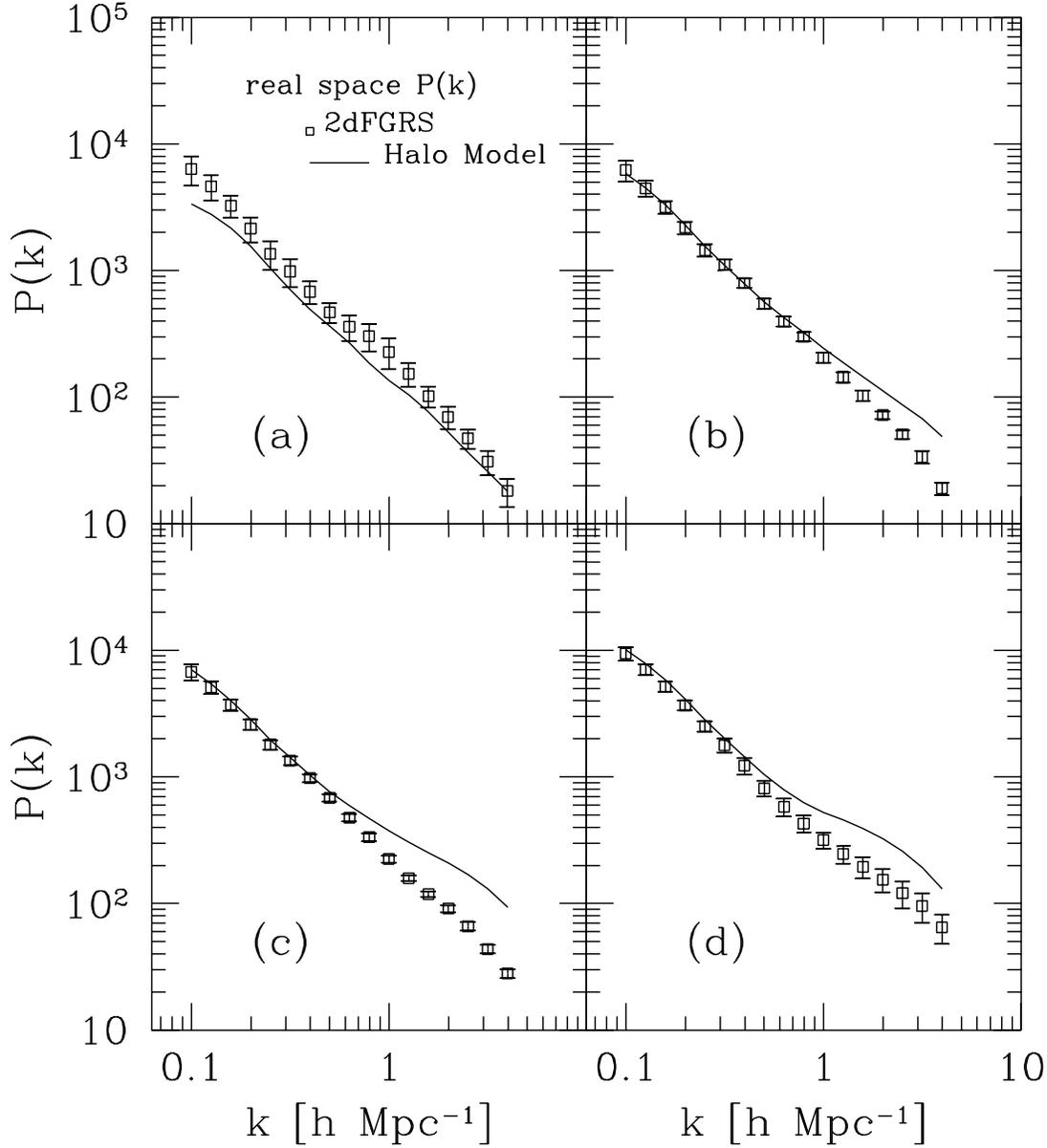}
\caption{The real space power spectrum $P(k)$ measured in the 2dFGRS
(symbols): a) for $-18.5< M_b-5\log_{10} h <-17.5$, b) for $-19.5<
M_b-5\log h <-18.5$, c) for $-20.5< M_b-5\log_{10} h <-19.5$, and d)
for $-21.5< M_b-5\log_{10} h <-20.5$. The error bars of the observed
results are given by the mock samples, as described in the text. The
smooth lines are the predictions based on the halo model. }
\label{fig:pkreal}\end{figure}

\begin{figure}
\epsscale{1.0} \plotone{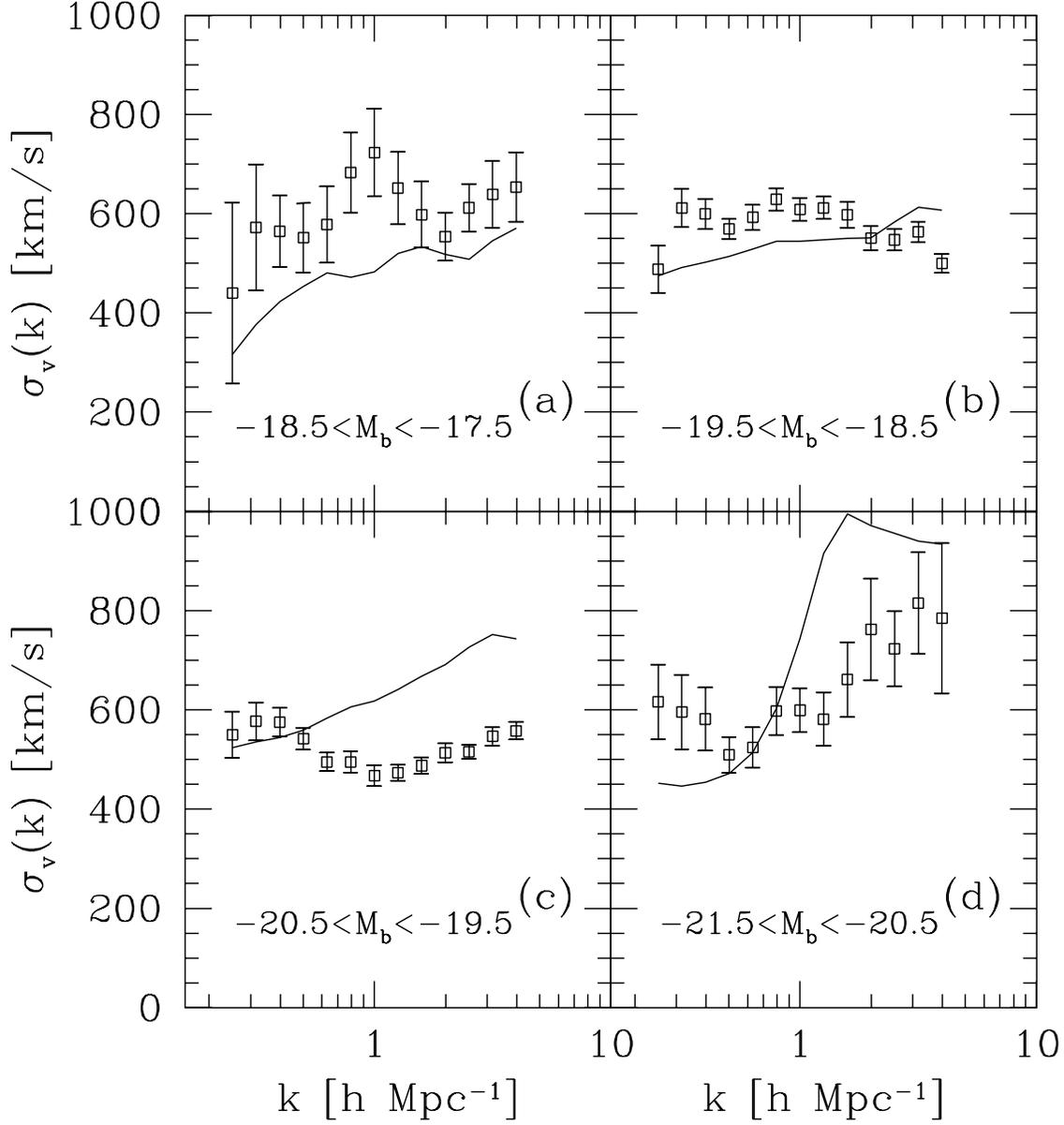}
\caption{The PVD of galaxies measured in the 2dFGRS (symbols),
compared with the predictions based on the halo model (solid
lines). The error bars of the observed results are given by the mock
samples, as described in the text.}
\label{fig:sigmav}\end{figure}

Our main results are shown in Figs.~(\ref{fig:pkmu}) to
(\ref{fig:sigmavk1}). In the introduction we have said that it should
be possible to derive $\beta$, $\sigma_{12}$, and $P(k)$ from a
measurement of the redshift space power spectrum
$P^S(k,\mu)$. However, because we measure $P^S(k,\mu)$ only up to the
scale $k=0.1\mpci$, there exists a strong degeneracy in determining
the parameters $\sigma_{12}$ and $\beta$ \citep{peacock00} from
$P^S(k,\mu)$. Moreover, $\sigma_{12}$ could be a function of
$k$\citep{jb01}. We therefore fix $\beta= 0.45$ as a reasonable
estimate \citep{hawkins03}, and then determine $P(k)$ and
$\sigma_{12}$ from the data. There could also be some luminosity
dependence in $\beta$, and we investigate this by using the luminosity
dependence of the bias parameter $b/b^* = 0.8 + 0.2 (L/L^*)$ given in
\citet[]{norberg02a} and $\beta^*=0.45$, where the quantities with
superscript $*$ are those at the characteristic luminosity $M^*$. The
result is shown in Fig.~(\ref{fig:sigmabetalum}). The PVD for the
brightest galaxies is slightly increased compared to the result in
Fig,~(\ref{fig:sigmavk1}), and it is slightly decreased for the faint
galaxies. The changes are small, therefore it seems justified to use a
constant $\beta$. In Figs.~(\ref{fig:sigmavk1}) and
(\ref{fig:sigmabetalum}) we have also plotted the PVD values at the
number-weighted luminosity value in each bin.

In Fig.~(\ref{fig:pkmu}) the basic measurement of the power spectrum
in redshift space $P^S(k, \mu)$ is shown. The four panels in this
figure correspond to four different luminosity intervals (samples
3,5,7, and 9 of table 1) from faint to bright galaxies. The symbols
are the results for the full 2dFGRS, the dotted lines for the south,
and the dashed lines for the north sample for each $k$-value. The
values of $k$ range from $0.20 \mpci$ at the top to $3.2 \mpci$ at the
bottom with an increment of $\Delta \log_{10}k = 0.2$.  The south and
north samples agree quite well with the full survey indicating that
cosmic variance is not a problem. The solid lines are the best
fits obtained by applying equation (\ref{eq2}) to the data of the
whole survey. The power spectrum for the larger $k$-values decreases
quite strongly with $\mu$, more than a factor of $10$ between $\mu =
0$ and $\mu = 1$. There is a small luminosity dependence. $P^S(k,\mu)$
increases for the bright samples by a factor $\simeq 2$ for all $k$. A
similar dependence on luminosity for the 2PCF was found by
\citet[]{norberg02a} (see also the real space spectrum below).

In Fig.~(\ref{fig:pkreal}) the real space power spectrum of the 2dFGRS
$P(k)$ is displayed for the four different luminosity intervals
(symbols). Again we can see that $P(k)$ for brighter samples are
higher.  The error bars are derived from the mock samples described in
the previous section. We prefer this error estimate to the bootstrap
method which in general gives $50 \%$ smaller errors. The mock sample
errors are just the standard deviations occurring, when the identical
analysis is carried out for the 20 mock samples. These error bars
adequately include the sample to sample variations (the cosmic
variance), and they are not sensitive to the bias model used (JMB98).
The figure shows that $P(k)$ is quite reliably determined for $k$
between $0.1 \mpci $ and $4 \mpci $. Although $P(k) = P^S(k,\mu =0)$
in principle, $P(k,\mu =0)$ fluctuates around the true value of $P(k)$
in practice, because in a finite sized survey the number of Fourier
modes is always limited. $P(k)$ is better determined, if $P^S(k,\mu)$
at different angles is combined with $P(k)$ being treated as a free
parameter, as is done here. The spectrum is approximately a power law
for the range of $k$ considered here. It decays with $k$ approximately
as $ k^{-1.5}$, with the slope of the brightest sample somewhat
shallower.  The smooth solid lines are the predictions of the halo
model of Yang et al. which is implemented as described in section
4. The agreement with the data is satisfactory on large scales where
$k<0.5\mpci$, not a surprising fact, since the halo model has been
constructed such as to reproduce the luminosity-dependent clustering
length of the 2dFGRS (i.e. linear to quasi-linear
regimes). Nevertheless, we may note that for samples 5,7, and 9 the
halo model gives an indication of a change in slope of $P(k)$ at $ k =
1 \mpci $. This can be understood as the transition between the scales
where the pair counts are dominated by galaxy pairs in the same halo
to the larger scales, where pairs of galaxies mostly are in separate
halos.  In the data such a change in slope is much less pronounced,
but it is present for the brightest samples \citep[see also
Y04,][]{zehavi03}. The discrepancy between the model and the 2dFGRS
data at small scales of $k\ga 1\mpci$ is very significant, at $\sim
3\sigma$ level for sample 9, at $\sim 10\sigma$ level for sample 7,
and at $\sim 5\sigma$ level for sample 5. We also note that the
clustering of the faint sample (sample 3) is higher than the halo
model on all scales at $\sim 1.5\sigma$ level.

The PVD is measured simultaneously with $P(k)$, and the results are
presented in Fig.~(\ref{fig:sigmav}). It seems that for the $k$-values
used here, $\sigma_{12}(k)$ is a well-determined quantity. For the
faintest sample, the PVD is persistently higher than that of the halo
model (at $\sim 2 \sigma$ level), indicating that the faint galaxies
in the Universe may reside in more massive halos than the halo model
predicts. This can also explain why the $P(k)$ of this sample is
systematically higher than the halo model prediction. For sample 5,
the difference in the PVD between the 2dFGRS and the halo model is
only $50$ to $100\kms$, consistent with the fact that their $P(k)$ are
also close, though the agreement is not very good as the discrepancy
between the model and the observation is significant at $2\sigma$
level. For the bright galaxies (sample 7 and sample 9), the halo model
prediction rises quickly with $k$ at $k>1 \mpci$, as the galaxy pairs
on these scales are mainly contributed by those from the same massive
halos. This behavior was also seen in the power spectrum $P(k)$ at
small scale where $P(k)$ rises quickly with $k$. Especially for sample
7, the observed PVD does not rise with $k$ and is only around
$500\kms$, which is expected if the galaxies are more distributed
among isolated halos with the velocity dispersion of $200\sim 300\kms$
(bright spirals). For sample 9, the PVD rises with $k$, but much
slower than the halo model around $1\mpci$, which also indicates that
the brightest galaxies in the real Universe are distributed more in
halos of different mass than the halo model predicts. The difference
between the halo model and the observation is significant at $>
3\sigma$ level for the bright samples.

The luminosity dependence of the PVD is shown most clearly in
Fig.~(\ref{fig:sigmavk1}), where we have plotted $\sigma_{12}$ at $ k=
1 \mpci $ for the 10 overlapping samples listed in table 1. The
surprising result is a strong dependence on luminosity with the bright
and the faint galaxies reaching high values of $ \simeq 700 \kms$, and
with a well defined minimum of $ \simeq 400 \kms$ for the galaxies of
sample 8. The bright and the faint galaxies apparently have high
random motions, as expected for objects in a massive halo or in a
cluster. The $M^*$ like galaxies are rather moderate in their PVD, and
probably reside in galaxy size halos.  The thick solid line represents
the prediction based on the halo model of Y04 (\S 4) that clearly does
not match our observation of the PVD.

\subsection{Implications for the halo model}
The results of Y04 as well as of the current work show that the halo
model adopted in this paper can reproduce the luminosity function very
well, but cannot match the clustering data of galaxies perfectly. The
model can quite successfully account for the luminosity dependence of
clustering on large scales (i.e. the luminosity dependence of the
clustering length), though it seems to underpredict for the faintest
galaxies (Figure 6). Moreover, the clustering at small scales
($r<2\mpc$ or $k>0.5\mpci$) predicted by the model is higher than the
2dFGRS data by a factor of 2 to 3. Also the PVD and the
quadrupole-to-monopole ratio of the redshift space correlation
function are higher in the model than the 2dFGRS data when the
luminosity classification has not been applied. To solve these problems
(or part of them), Y04 proposed several possible solutions, that is,
to raise the mass-to-light ratio to $1000(M/L)_\odot$ for rich
clusters in the B-band, to introduce a strong velocity bias $b_{\rm
vel}\equiv \sigma_{\rm gal}/\sigma_{\rm dm}=0.6$ (where $\sigma_{\rm
gal}$ and $/\sigma_{\rm dm}$ are the velocity dispersions of galaxies
and dark matter in halos), or to lower the clustering amplitude
$\sigma_8$ to about $0.7$. As Y04 realized, the first two solutions
are not realistic, because the observed mass-to-light ratio is
$(450\pm 100)(M/L)_\odot$ \citep{fuku98} and $(363\pm 65)(M/L)_\odot$
\citep{cal96} for rich clusters in the B-band, and hydro/N-body
simulations show that velocity bias is rather minor for rich galaxy
systems \citep[][]{yoshikawa03,berlind03}. Therefore we only consider
the last possibility to see if a low $\sigma_8=0.7$ can bring the halo
model into agreement with our luminosity-dependent PVD result.

In our N-body data library, we have one simulation with $L=100\mpc$
and another with $L=300\mpc$ in a $\sigma_8=0.7$ LCDM model. The model
and simulation parameters of these two simulations are the same as
those of the simulations with the same boxsize $L$ in \S 4, except for
that the density parameter $\Omega_0=0.25$ is slightly lower for the
new model. Xiaohu Yang has kindly provided us with the parameters of
the halo model for this cosmological model. We find that the
luminosity function of the 2dFGRS can be reproduced and the clustering
is improved on small scales (but still higher for luminosity bins
around $M^*$) in this model. Its PVD is shown in Fig.8 as a function
of luminosity, which is considerably lower than the data of
2dFGRS. Compared with the low $\sigma_8$ model of Y04, our $\Omega_0$
is slightly lower. Since the PVD is approximately proportional to
$\sigma_8 \Omega_0^{0.6}$, we can boost the PVD by $(0.30/0.25)^{0.6}$
which is also shown in Fig.8. We can see that this PVD after
correcting for the $\Omega_0$ difference is still lower than our
2dFGRS data. More interestingly, this low $\sigma_8$ model also predicts
a monotonically decreasing PVD with the decrease of the luminosity,
implying that simply changing the value of $\sigma_8$ cannot overcome
the difficulty of the halo model to explain the bimodal nature of the
luminosity dependence of the PVD.

It should be possible to use the PVD to constrain the value of
$\sigma_8$ if the shape of the PVD is reproduced in the halo
model. Y04 argued for $\sigma_8=0.7$ from a comparison with the 2dFGRS
PVD without a luminosity classification \citep[][]{hawkins03}
\citep[see also][for other independent arguments]{bosch03}, but our
comparison of the halo model with our 2dFGRS data seems to favor a
higher $\sigma_8$, since the PVD of the this low $\sigma_8$ model is
lower than the observed data at all luminosities. Considering that
most of the 2dFGRS galaxies are in the luminosity interval
$-20.5<\mag<-18.5$, we would require $\sigma_8\approx 0.82$ for
the $\Omega_0=0.3$ according to the $\sigma_8\Omega_0^{0.6}$ scaling
if we want the halo model to match the observed PVD at
$\mag=-19.5$. This value of $\sigma_8$ is in good agreement with the
recent determinations based on the WMAP and the SDSS galaxy-galaxy lensing
data \citep[][]{spergel03,seljak04a}. Although this is not a rigorous
determination for $\sigma_8$ because we have not yet reproduced the
shape of the luminosity dependence of PVD, our comparison indicates
that the concordance model favored by the WMAP and SDSS data is
consistent with our PVD data.

The PVD is an indicator of the depth of the local gravitational
potential. Therefore we are inevitably led to the interesting
conclusion that the bright and the faint galaxies move in the
strongest gravitational field. A substantial fraction of them must be
in clusters, while most galaxies around magnitude $\mag =
-20.5$, the $M^*$ galaxies in the Schechter luminosity function,
populate the field. We believe this result, that is, the shape of the
PVD luminosity dependence (not its absolute amplitude),
constitutes a new challenge to the halo model of Yang et al.. The
straightforward implication for the halo model is to increase the
faint galaxy population in rich clusters. With the parameters of the
Yang et al.'s halo model, the faint end slope $\alpha_{15}$ of their
conditional luminosity function in rich clusters is one of the crucial
quantities that determine the fraction of the faint galaxies in rich
clusters. We expect that decreasing $\alpha_{15}$ (i.e. steeper faint
end slope) can increase the fraction of faint galaxies in rich
clusters. It is therefore worth investigating if a more negative
$\alpha_{15}$ than the value $-1.32$ of the M1 model can save the halo
model of Y03. 

It is important to emphasize that the conditional luminosity function
method proposed by Yang et al. is a highly valuable approach to infer
the galaxy distribution within dark halos from a wide range of
observations. The discrepancy between the M1 model and our PVD
luminosity dependence (especially the bimodal shape of the PVD) does
not necessarily mean that the halo model approach is wrong or
useless. Instead it does mean that the luminosity dependence of the
PVD provides an important test for galaxy formation models that is
independent of the correlation functions. Because the PVD is more
sensitive to the small number of galaxies in rich clusters than the
correlation function \citep{mjb93,mjb97}, combining these two types of
statistical quantities can put a more stringent constraint on the
parametrization of the halo model.

The galaxy-galaxy lensing of the SDSS constrains strongly the fraction
of faint galaxies in rich clusters \citep{sheldon04,mckay02}. The
rapid decrease of the lensing signal with the luminosity of the
lensing galaxies indicates that the average matter density around
galaxies decreases with the decrease of galaxy luminosity
\citep{primack04,seljak04b}. However, the galaxy-galaxy lensing
observation may not be inconsistent with our PVD observation that
quite a fraction of faint galaxies are in rich clusters. The main
reason is that these two quantities as well as the two-point
correlation function depend on the halo occupation number of galaxies
differently. In other word, the faint population in rich clusters has
a different weight on these different quantities. Because of its
pair-square weighting nature, the PVD statistic is mostly sensitive to
those faint galaxies in rich clusters \citep{mjb97}. Because the mean
lensing signal \citep{sheldon04} is galaxy number weighted, the
lensing statistic may be much less sensitive to the faint population
in rich clusters than the PVD. Figure 6 also shows that the two-point
correlation predicted by the current halo model (M1 model) is about 50
percent lower than the observed value for faint galaxies, indicating
there is room for increasing faint galaxies in rich clusters.  It would
certainly be very interesting to examine quantitatively if such a halo
model can be constructed to consistently interpret all these three
quantities. Because these quantities depend on the halo occupation
number in different manners, we expect that they will play important
but complementary roles in constraining the halo occupation of
galaxies.

\section{Conclusions}

The analysis of the velocity fields of the galaxies in the 2dFGRS
(C01) has led to a surprising discovery: The random velocities of the
faint galaxies are very high, around $ 700 \kms$, reaching similar
values as the bright galaxies. At intermediate luminosities the
velocities exhibit a well defined steep minimum near $ 400 \kms$.

\begin{figure}
\epsscale{1.0} \plotone{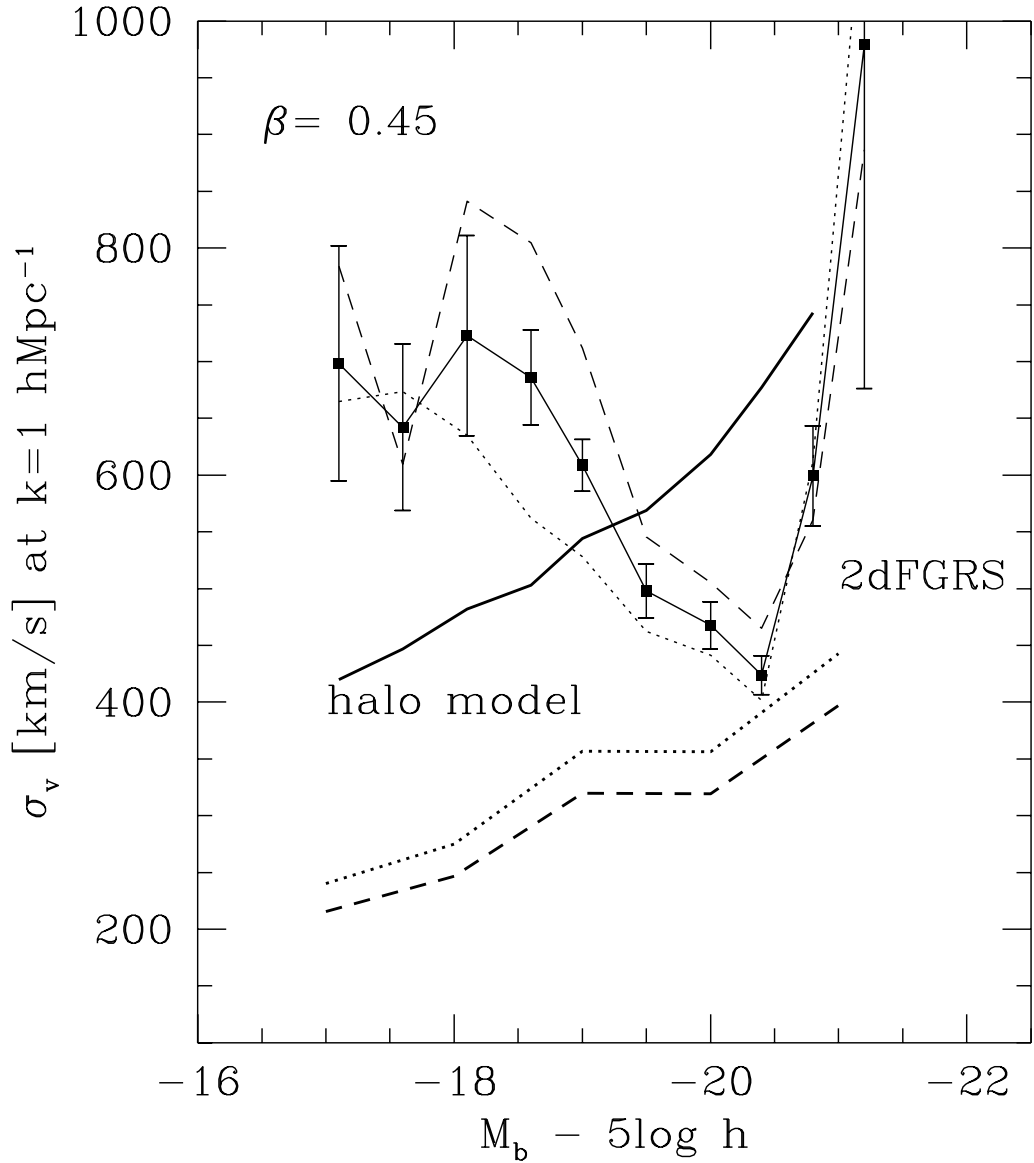}
\caption{The PVD measured at $k=1\mpci$ in the 2dFGRS (symbols for the
whole sample, dotted line for the south, and dashed line for the
north), compared with the predictions based on the halo model. The
thick solid line is for the nominal model of $\sigma_8=0.9$, the thick
dashed one is for the model of $\sigma_8=0.7$, and the thick dotted
one is for the model of $\sigma_8=0.7$ but with $\Omega_0=0.3$. The
error bars of the observed results are given by the mock samples, as
described in the text.}
\label{fig:sigmavk1}\end{figure}

\begin{figure}
\epsscale{1.0} \plotone{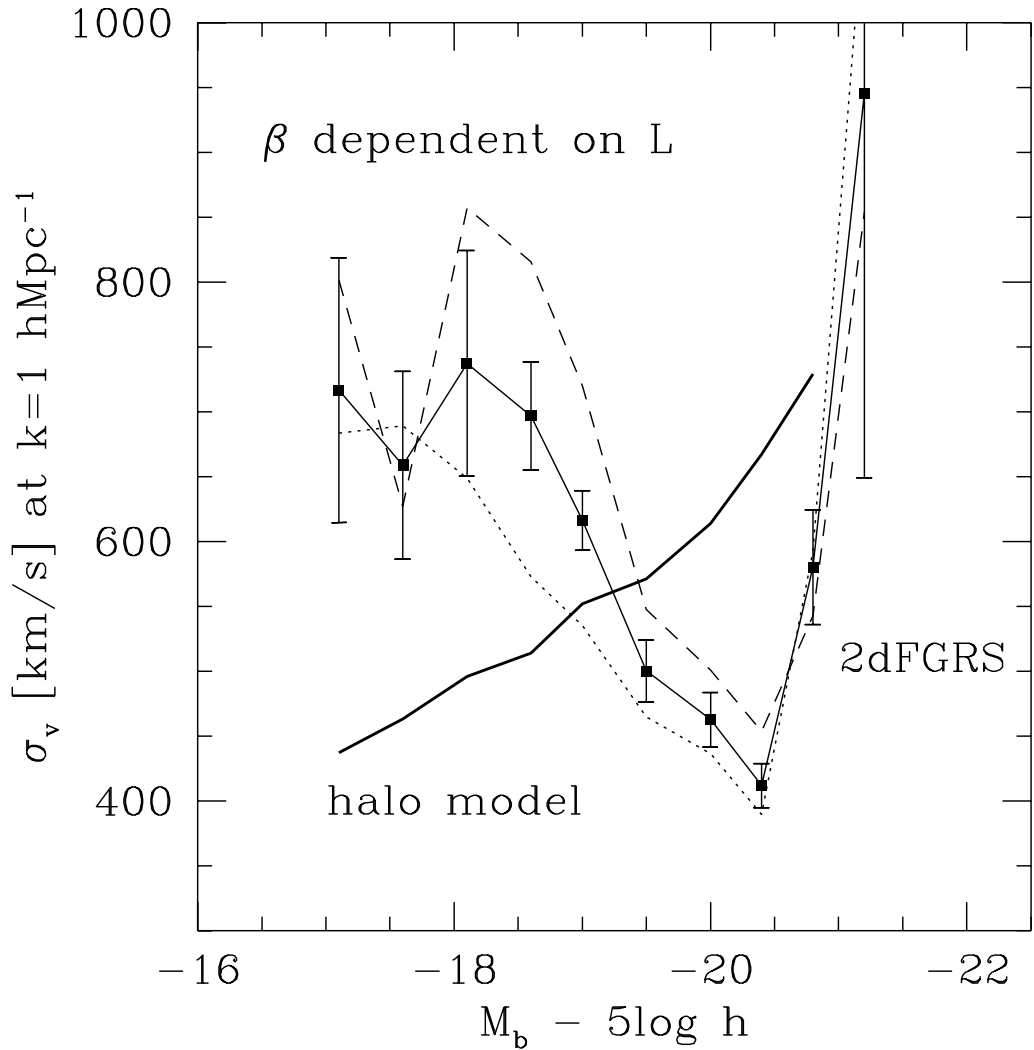}
\caption{The PVD measured at $k=1 \mpci$ in the
2dFGRS. Symbols have the same meaning as in Fig.~(\ref{fig:sigmavk1}).
Here the parameter $\beta$ varies with luminosity as in \citet[]{norberg02a}.
}
\label{fig:sigmabetalum}\end{figure}

It seems that the galaxies in different luminosity intervals appear as
different populations in their own right, defined by objective
statistics.  A look at Fig.~(\ref{fig:sigmavk1}) shows convincingly
that this is actually the case. For this figure we have sorted the
galaxies in 10 luminosity bins, each one magnitude wide, from
magnitude $-16.5$ to $ -22$ and plotted the value of $ \sigma_{12}$ at
a wave number of $k \simeq 1 \mpci $. Such a finely resolved binning
of galaxies in samples of different luminosities is possible for the
2dFGRS redshift survey, because it is big enough to contain
sufficiently many galaxies in each luminosity class. In each bin the
PVD is a well defined quantity which can be measured reliably ( the
very luminous galaxies have large error bars, because their are very
few pairs of such objects at those scales).

The PVD is an indicator of the depth of the local gravitational
potential. Therefore we find the interesting result that the bright
and the faint galaxies move in the strongest gravitational field. A
substantial fraction of them must be in clusters, while most galaxies
around magnitude $\mag = -20.5$, the $M^*$ galaxies in the
Schechter luminosity function, populate the field.

The bimodal nature of the correlation between PVD and luminosity may
be used as a stringent test of galaxy formation models.  We have
investigated the halo occupation model \citep[Y03][]{} which has been
optimally fitted to reproduce the luminosity function, and the
two-point correlation function of the 2dFGRS. If we adapt this model
to the PVD value of the $M^*$ galaxies, we see that it cannot give the
high values found for the fainter galaxies. The PVD values of the
model actually run opposite to the data and the model assigns smaller
values to the fainter galaxies.  This indicates that the assignment of
galaxies to the dark matter halos must be done in a more intricate way
as up to now. In Table 2 we have listed the PVD values for the
different luminosity bins. These must be reproduced by an acceptable
model for galaxy formation.  The number of faint galaxies in clusters
must be increased substantially to at least recover the high PVD found
for them.  One possible solution for the halo model is to raise the
faint end slope of the conditional luminosity function in rich
clusters. Also, the low value of $ 400 \kms$ found for the galaxies
with magnitude $ -20.5$ must mean that these galaxies reside in dark
matter halos of galactic size,

Another way, widely used, to connect dark matter to galaxies is the
semi-analytic modeling \citep[e.g., ][]{kauffmann99, cole00, benson00,
sp99, dkcw99,springel01}, where the dark matter distributions obtained
from N-body simulations are supplemented with some of the physical
processes important in galaxy formation using semi-analytic
techniques. A test of the PVD vs luminosity for this type of models
will be the aim of a subsequent paper.

We have shown here in addition that our novel method of deriving
the real space power spectrum and the PVD from the redshift space
power spectrum allows to determine these quantities precisely
and reliably from the 2dFGRS. The size of the 2dFGRS permits 
an investigation of the luminosity dependence of these quantities.
New constraints on galaxy formation models can be derived from that.
We have shown that the luminosity dependence of the clustering
can be reproduced by an adequately chosen halo occupation model.
This model fails, however, to reproduce the luminosity dependence of
the PVD.

\acknowledgments 

We are grateful to Houjun Mo for a very stimulating and helpful
discussion on the halo model, Uros Seljak for his helpful comment on
the galaxy-galaxy lensing observation, and Xiaohu Yang for providing
the halo model parameters of the low $\sigma_8$ model and for some
instructions of implementing the halo model in N-body simulations, and
the anonymous referee for helpful comments that improve our
presentation.  JYP would like to thank the Max-Planck Institute f\"ur
Astrophysik for its warm hospitality during the time when this work
was completed.  GB wants to thank the Shanghai Astronomical
Observatory for friendly infrastructure provided during his stay in
Shanghai. The work is supported in part by NKBRSF (G19990754), by NSFC
(Nos.10125314, 10373012), and by the CAS-MPG exchange program.

\clearpage
 
\begin{table}
\caption{Samples selected according to luminosity}
\begin{center}
\begin{tabular}{cccccccc}
\hline\hline 
&&South&North&Total\\
Sample&$M_b-5\log_{10} h$&(no. of galaxies)&(no. of galaxies)&(no. of galaxies)\\
\hline 
1&$-16.5>M_b-5\log_{10} h\ge -17.5$&5218&   3587&  8805\\ 
2&$-17.0>M_b-5\log_{10} h\ge -18.0$&9314&   6078& 15392\\ 
3&$-17.5>M_b-5\log_{10} h\ge -18.5$&14593& 11097& 25690\\ 
4&$-18.0>M_b-5\log_{10} h\ge -19.0$&23703& 18670& 42373\\ 
5&$-18.5>M_b-5\log_{10} h\ge -19.5$&34481& 25683& 60164\\ 
6&$-19.0>M_b-5\log_{10} h\ge -20.0$&40995& 29241& 70236\\ 
7&$-19.5>M_b-5\log_{10} h\ge -20.5$&40182& 27223& 67405\\ 
8&$-20.0>M_b-5\log_{10} h\ge -21.0$&30934& 19204& 50138\\ 
9&$-20.5>M_b-5\log_{10} h\ge -21.5$&15388&  8848& 24236\\ 
10&$-21.0>M_b-5\log_{10} h\ge -22.0$&3739&  2162& 5901 \\ 

\hline\hline
\end{tabular}
\end{center}
\end{table}
\begin{table}
\caption{The results of the PVD at $k=1\mpci$ of the luminosity subsamples}
\begin{center}
\begin{tabular}{cccccccc}
\hline\hline 
&Median Mag.&$\sigma_v[\beta=0.45]$\tablenotemark{a}&$\sigma_v[\beta(L)]$\tablenotemark{b}\\
Sample&$M_b-5\log_{10} h$&($\kms$)&($\kms$)\\
\hline 
1  & $-17.15$ & $698\pm 103$ & $717 \pm 102$\\ 
2  & $-17.62$ & $642\pm  73$ & $659 \pm  72$\\ 
3  & $-18.12$ & $723\pm  88$ & $737 \pm  87$\\ 
4  & $-18.61$ & $686\pm  42$ & $697 \pm  42$\\ 
5  & $-19.06$ & $609\pm  23$ & $616 \pm  23$\\ 
6  & $-19.51$ & $498\pm  24$ & $500 \pm  24$\\ 
7  & $-19.96$ & $468\pm  21$ & $463 \pm  21$\\ 
8  & $-20.39$ & $424\pm  17$ & $412 \pm  17$\\ 
9  & $-20.76$ & $599\pm  44$ & $580 \pm  44$\\ 
10  & $-21.18$ & $979\pm 303$ & $945 \pm 296$\\ 

\hline\hline
\end{tabular}
\end{center}
\small{
\tablenotetext{a}{The PVD determined with $\beta=0.45$.}
\tablenotetext{b}{The PVD determined with the luminosity dependence of $\beta$ taken into account.}}
\end{table}

\end{document}